\documentclass[11pt]{article}
\usepackage{psfig,amsmath,amssymb,afterpage}
\newenvironment{mydescription}[1]%
 {\begin{list}{}{%
  \setlength{\leftmargin}{15pt}%
  \setlength{\itemindent}{15pt}%
  \setlength{\rightmargin}{15pt}}}
  {\end{list}}

\def\lsi{\raise0.3ex\hbox{$<$\kern-0.75em\raise-1.1ex\hbox{$\sim$}}}
\def\gsi{\raise0.3ex\hbox{$>$\kern-0.75em\raise-1.1ex\hbox{$\sim$}}}
\newcommand{\lsim}{\mathop{\lsi}}
\newcommand{\gsim}{\mathop{\gsi}}

\begin{document}

\begin{flushright}
 BI-TP 98/31 \\
 September 1998
\end{flushright}

\begin{center}

\vspace{24pt}
{\LARGE \bf   Lattice Gravity and Random Surfaces }
\vspace{24pt}

{\Large \sl Gudmar Thorleifsson} \\
\vspace{10pt}
Facult\"{a}t f\"{u}r Physik, Universit\"{a}t Bielefeld
   D-33615, Bielefeld, Germany  \\
\vspace{10pt}

\begin{abstract}
I review recent progress in simplicial quantum gravity in
three and four dimensions, in particular new results on the
phase structure of modified models of dynamical triangulations,
the application of a strong-coupling expansion,
and the benefits provided by including degenerate triangulations. 
In addition, I describe some
recent numerical and analytical results on 
anisotropic crystalline membranes.
\end{abstract}

\end{center}


\section{Simplicial Quantum Gravity in $\mathbf D > 2$}

\noindent
The task of formulating a consistent theory of
quantum gravity in four dimensions, hence unifying
the two pillars of modern physics, general relativity
and quantum mechanics, is a formidable one and
still unresolved.  Many different approaches have been 
tried (see e.g.\ Ref.~\cite{gen}), some have
led to a deeper understanding of the problem
but, as of yet, none can claim much success.
In this article I will review the status, and recent
progress, of one such approach, namely the attempt to 
make sense of an Euclidean path-integral quantization 
of general relativity via a discretization known
as dynamical triangulations.

In Euclidean quantum gravity a path-integral is written
as a formal sum over $D$--dimensional Euclidean geometries
(metrics $g_{\mu \nu}$)
\begin{equation}
 Z \;=\; \int \frac{{\cal D}[g]}{\rm Vol(diff)} \; 
  {\rm e}^{\textstyle - S_E[g_{\mu \nu}]}
 \label{contpart}
\end{equation}
weighted by the Einstein-Hilbert action
\begin{equation}
 S_E[g_{\mu \nu}] \;=\; \frac{1}{16 \pi G} \int
       {\rm d}^D\xi \; \sqrt{|g|} (-R + 2\Lambda).
 \label{EHaction}
\end{equation}
$G$ is the Newton's constant, $\Lambda$ the cosmological
constant and $R$ the scalar curvature. However, this
path-integral formulation faces 
some sever problems and unresolved questions:

\begin{mydescription}

\vspace{-8pt}
\item[$\bullet$]
 \hspace{-13pt} the theory is non-renormalizable,
 
\vspace{-8pt}
\item[$\bullet$]
 \hspace{-13pt} the action is unbounded from below, 
 
\vspace{-8pt}
\item[$\bullet$]
 \hspace{-14pt} adding higher-order curvature terms 
 leads to renormalizable, albeit non-unitary, theory, 

\vspace{-8pt}
\item[$\bullet$] 
 \hspace{-12pt} how can Lorentzian signature be recovered from the
 Euclidean formulation.
 
\end{mydescription} 

\vspace{-3pt}
A proposed interpretation is to view Eq.~(\ref{EHaction})
as the first part of an effective action, valid in the infrared
limit, and that fine-tuning of the parameters in
the full effective theory might reveal a non-trivial 
{\it ultraviolet} stable fixed point where a consistent quantum theory
of gravity might be defined \cite{wein}  
(a behavior akin to that of the non-linear $\sigma$-model).
While it is true that non-trivial fixed points in four-dimensional field
theories are rare, some interesting results obtained in 2+$\epsilon$ 
dimensions actually lend support to this idea \cite{2eps}.  

To address this proposal, a non-perturbative investigation of
the theory is needed.  One such is provided by a discretization 
of the path-integral Eq.~(\ref{contpart}), usually
following a prescription due to Regge \cite{regge}. 
He proposed to approximate a smooth manifold with a given 
metric by a piecewise linear manifold
(a {\it triangulation}), on which both parallel
transport and the integral of curvature are well-defined.  
Choosing a sufficiently fine triangulation
should provide a good approximation to any continuum
manifold. 

This idea has been applied to lattice gravity 
following one of two complimentary prescriptions,
differing in how they implement the dynamical nature
of the space-time metric:

\begin{mydescription}

\vspace{-4pt}
\item[(a)]
{\it Quantum Regge calculus} \\
Use a piecewise linear manifold with {\it fixed} 
connectivity but with {\it varying}  edge lengths
\cite{xregge}.

\vspace{-4pt}
\item[(b)]
{\it Dynamical Triangulations} \\
Summing over triangulations with {\it varying} 
connectivity but {\it fixed} edge lengths~\cite{dyn}.
\end{mydescription}

\vspace{-4pt}
\noindent
In this review I will focus on the latter approach,
for a recent review of quantum Regge calculus
see e.g.\ Ref.~\cite{williams}. 

\subsection{Dynamical triangulations}

In models of dynamical triangulations the continuum integral
over diffeomorphism inequivalent metrics is replaced by
a discrete sum over all possible decompositions (gluings)
of equilateral $D$--simplexes along their ($D$$-1$)--dimensional
faces into simplicial manifolds, while
requering that the neighbourhood of each vertex is
a $D$--ball (excluding {\it pseudo-manifolds}) \cite{david,janbook}.
Fixing all edge lengths to a constant $a$ introduces a
``reparametization invariant'' cut-off in the model.

In this approach the Einstein-Hilbert action is approximated
by a simple function of the numbers $N_i$ of 
$i$--dimensional simplexes on a given triangulations $T$.
However, as these numbers are related through the 
Dehn-Sommerville and Eulers relations, the action
can be taken to depend on only two of those numbers.
The (grand-canonical) partition function becomes:
\begin{equation}
 Z(\mu,\kappa) = 
  \sum_{N_D} {\rm e}^{- \mu N_D} \hspace{-3pt} 
  \sum_{T \in {\cal T}_{N_D}}
   \frac{1}{C(T)} \, {\rm e}^{\,\kappa N_{D-2}},
 \label{part.simplicial}
\end{equation}
where $\mu$ and $\kappa$ represent the discrete cosmological
and inverse Newtons's constants, and $C(T)$ is the
order of the automorphishm group of $T$ --- the number of
equivalent labelings of the vertexes.

The second sum is over a suitable ensemble ${\cal T}$ of 
$D$--dimensional triangulations (of volume $N_D$).
A priori, different ensembles can be used provided
they lead to a well-defined partition function.
While this leaves ample freedom in defining different
ensembles, the topology of the triangulations
must be fixed else the partition function is 
divergent in $N_D$. So far,
most simulations in $D > 2$ have used triangulations
of spherical topology.

The main justification for the dynamical triangulation
approach comes from two-dimensions where variety of
exact solutions exist, complemented by extensive numerical 
results, and which agree with the solutions
of continuum Liouville theory.\footnote{In contrast, 
simulations of quantum Regge calculus in two dimensions
fail to reproduce the known continuum results \cite{regge2}.}

\subsection{Simulation methods}

The partition function Eq.~(\ref{part.simplicial}) is evaluated
numerically using a Monte Carlo algorithm.
The space of all triangulations is explored using
a set of $D$+1 local (topology preserving) geometric
moves --- the ($p\,$,$\,q$)--moves \cite{moves}.
In a ($p\,$,$\,q$)--move, where $p = D + 1 - q$,
  a ($q\,$$-$$1$)--subsimplex is replaced 
by its ``dual''   ($p\,$$-$$1$)--subsimplex, provided
no manifold constraint is violated.  The moves
are ergodic in $D \leq 4$, i.e.\ any triangulation
can be transformed into any other by an appropriate
sequence of moves.  
An example of the ($p\,$,$\,q$)--moves in three dimensions is shown in 
n~1.

\begin{figure}[t]
 \centerline{\psfig{figure=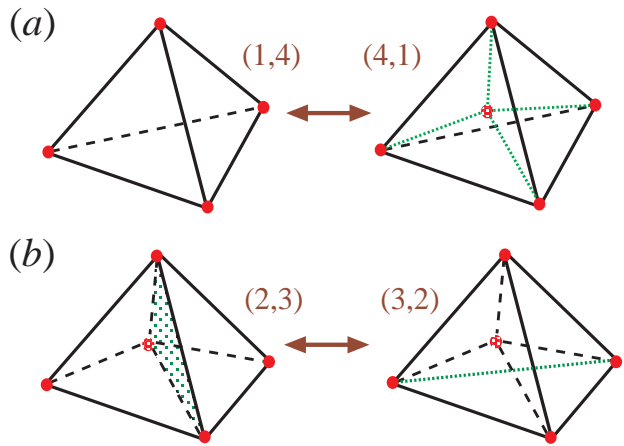,width=3.2in}}
 \label{fig:move3d}
 \caption{\small The ($p\,$,$\,q$)--moves in three dimensions: ({\it a})
  inserting/deleting a vertex, ({\it b}) replacing a triangle
  by an edge and vice verse.}
\end{figure}

In $D > 2$ fluctuations in the volume $N_D$ are
necessary for an ergodic updating procedure
and it is customary to simulate the quasi-canonical 
partition function:

\begin{equation}
 Z_{\bar{N}_D}(\mu,\kappa) 
  \;=\; \sum_{N_D} {\rm e}^{- \mu N_D}
 \sum_{T \in {\cal T}_{N_D}}
   \frac{1}{C(T)} \, {\rm e}^{
    \, \kappa N_{D-2} -  \delta
      (N_D - {\bar{N}_D)^2} }, 
 \label{part.quasi}
\end{equation}
where the volume fluctuates around a
target volume $\bar{N}_D$ and the quadratic potential
term added to the action ensures that, 
for an appropriate choice of  $\delta$, 
the fluctuations are small.

To explore the phase structure of model Eq.~(\ref{part.quasi})
powerful methods for probing the nature of the quantum geometry 
have been developed.  The fractal structure 
of the triangulations is labeled by a set of critical 
exponents, e.g:

\vspace{5pt}

\noindent
({\tt i}) The {\it string susceptibility exponent} 
$\gamma_s$ defines the singular behavior of 
the grand-canonical partition function:
$Z(\mu) \approx Z_{\rm reg} + (\mu - \mu_c)^{2-\gamma_s}$.
This in turn implies that the canonical partition function
behaves asymptotically as
\begin{equation}
 Z_{N_D}(\kappa) \;\sim\; {\rm e}^{\; \mu(\kappa) N_D}
 \;N_D^{\gamma_s(\kappa)-3}, 
 \;\;\;\; N_D \rightarrow \infty.
 \label{partasymp}
\end{equation}
A powerful method for measuring $\gamma_s$ in quasi-canonical
simulations is provided by the size distribution of 
{\it minbu's} (minimal neck baby universes) \cite{minbu}
--- a part of the triangulation connected to the
rest {\it via} a minimal neck.  By counting in how many ways
a minbu of size $B$ can be connected to a surface of size
($N_D-B$), the size distribution becomes
\vspace{-2pt}
\begin{eqnarray}
 n_{N_D}(B) &\approx 
   &\frac{B\,Z_B\,(N_D-B)\,Z_{N_D-B}}{Z_{N_D}}  \\ \nonumber
   &\sim &\left [ (N_D - B)\,B \right ]^{\gamma_s - 2}.
 \label{eq.baby}
\end{eqnarray}
The distribution $n_{N_D}(B)$ is measured in
simulations and $\gamma_s$ 
determined by a fit to Eq.~(6).

\vspace{5pt}

\noindent
({\tt ii}) The {\it fractal} or {\it Hausdorff dimension} $d_H$
measures the intrinsic ``dimensionality'' of the
triangulations and is defined by the volume of a geodesic
ball with radius $r$: $v(r) \sim r^{d_H}$.  
The geodesic distance is defined as the shortest path $d_{ij}$,
between two vertexes, traversed along links, either on the 
triangulations or its dual graph.

To measure the fractal dimension $d_H$ one explores the
scaling behavior of the vertex-vertex, or two-point,
correlation function,
\begin{equation}
 g_{N_D}(r) \;=\; \frac{1}{N_D} \; 
  \langle\;\sum_{i,j} \delta(d_{ij}-r)\,\rangle_T,
 \label{vertex}
\end{equation}
which counts the number of vertexes at distance
$r$ from a marked vertex $i$.  Assuming that the only relevant 
length-scale in the model is defined by $N_D^{1/d_H}$,
general scaling arguments \cite{dhscal1} imply that
\begin{equation}
  g_{N_D}(r) \;\sim\; N_D^{1-1/d_H}\,F
  \left (\frac{r}{\textstyle N_D^{1/d_H}}  \right ).
 \label{vscal}
\end{equation}
The optimal scaling of distributions $g_{N_D}(r)$, 
corresponding to different volumes,
onto a single scaling curve defines $d_H$.

\subsection{The pure gravity phase structure}

Extensive numerical simulations have established
that the model Eq.~(\ref{part.quasi}) has 
two distinct phases,
both in three and four dimensions.
There is a strong-coupling (small $\kappa$)
{\it crumpled} phase and a weak-coupling (large $\kappa$) 
{\it elongated} phase.  The two phases
are separated by a phase transition at critical value 
of the inverse Newton's constant~$\kappa_c$.

\begin{figure}[t]
 \centerline{\psfig{figure=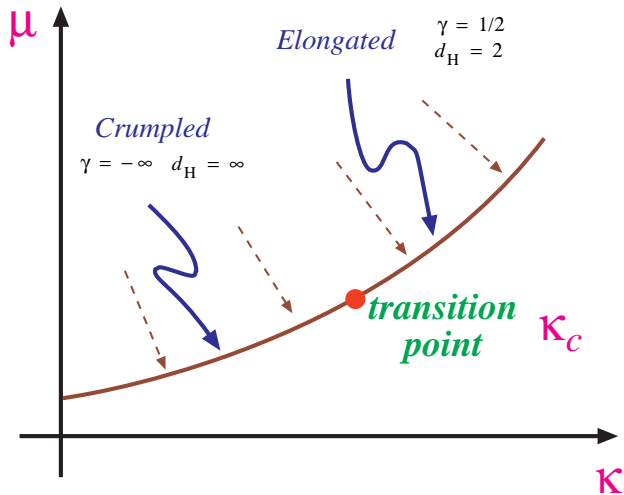,width=3.2in}}
 \label{fig.pgphase}
 \caption{\small A schematic phase diagram of pure simplicial
  gravity in $D > 2$.}
\end{figure}

For $\kappa < \kappa_c$ the internal structure collapses
and the crumpled phase is characterized by 
a {\it singular structure}, i.e.\ 
a small set of sub-simplexes connected to an extensive
fraction of the total volume \cite{singular}.
In $D=4$ this structure consists
of {\it two} singular vertexes, with local volumes $\omega^{\prime}_v$ 
that grows linearly with the volume $N_4$, joined
by a sub-singular edge;
$\omega^{\prime}_e \sim N_4^{2/3}$.  
In $D=3$ there is only a sub-singular vertex with 
$\omega^{\prime}_v \sim N_3^{1/3}$.
The geodesic distance between any two vertexes
always stays at the level of the lattice cut-off.
This prevents any sensible continuum
limit in this phase --- the quantum geometry is
essentially zero-dimensional.  In addition,
large geometric out-growths are suppressed and, at least 
formally\footnote{In the crumpled phase 
the sub-leading asymptotic behavior is
$Z_{N_D}(\mu) \sim \exp(\mu N_D + a N_D^b)$, $b<1$,
rather than Eq.~(\ref{partasymp}).
Hence $\gamma_s$ is not well defined.},
$\gamma_s = - \infty$.

In the elongated phase the internal geometry is
composed of bubbles glued together {\it via} small necks
into tree-like structures --- {\it branched polymers} ---
with $d_H = 2$ and $\gamma_s = 1/2$.
The neck size does not grow with the volume
and the internal geometry is essentially one-dimensional.

As the fractal structures described above are
somewhat pathological, it is unlikely that an interesting
continuum limit exist in either of the two phases.
The initial hope was, however, that the transition point
$\kappa_c$ would correspond to a non-trivial 
fixed point where
a sensible continuum limit could be defined. 
However, the transition turned out to
be {\it discontinuous}, both in three \cite{old3d,jap3d} 
and four \cite{pet4d} dimensions; this is
demonstrated in Figure~3.

\begin{figure}[t]
 \centerline{\psfig{figure=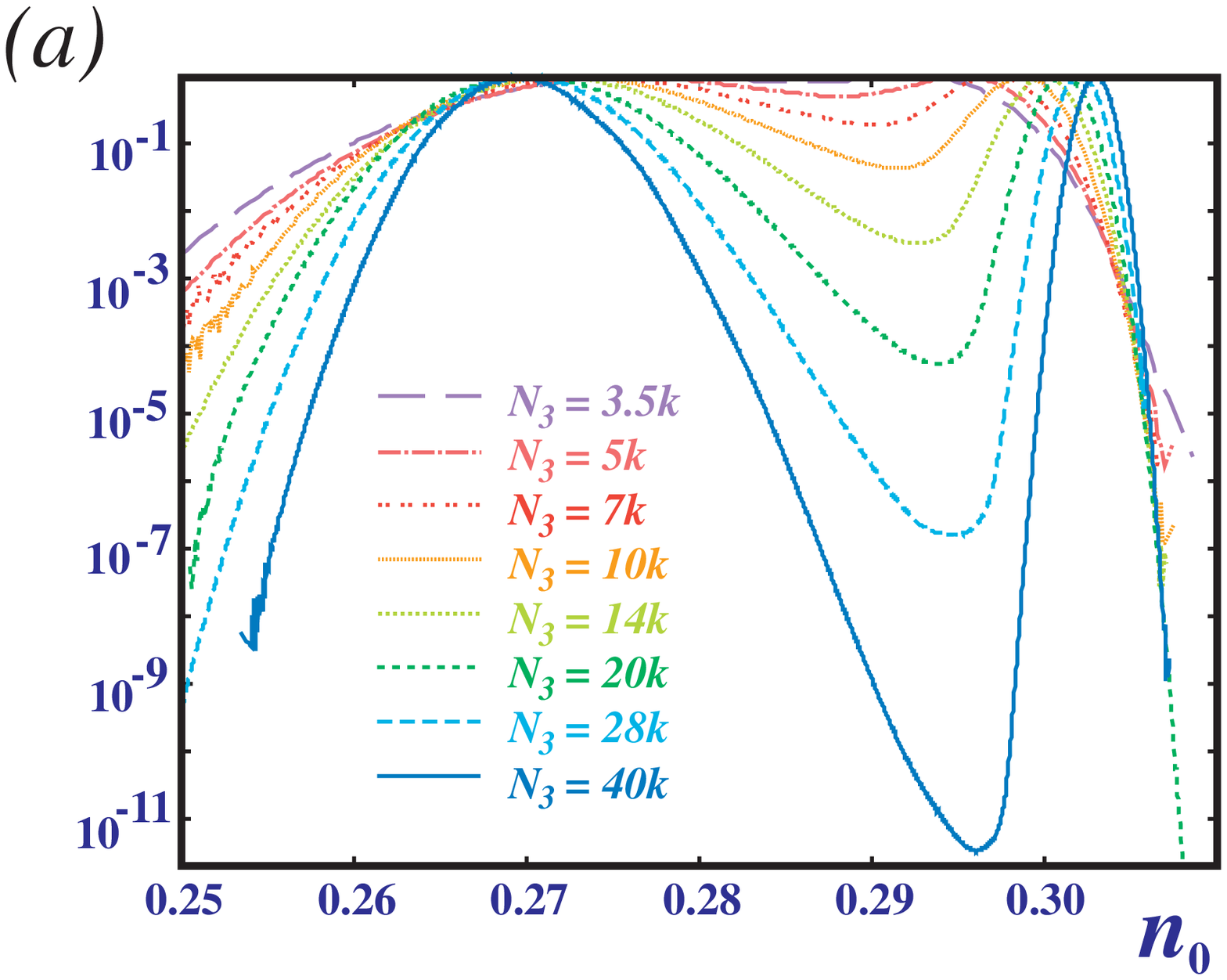,width=3.2in}}
 \centerline{\psfig{figure=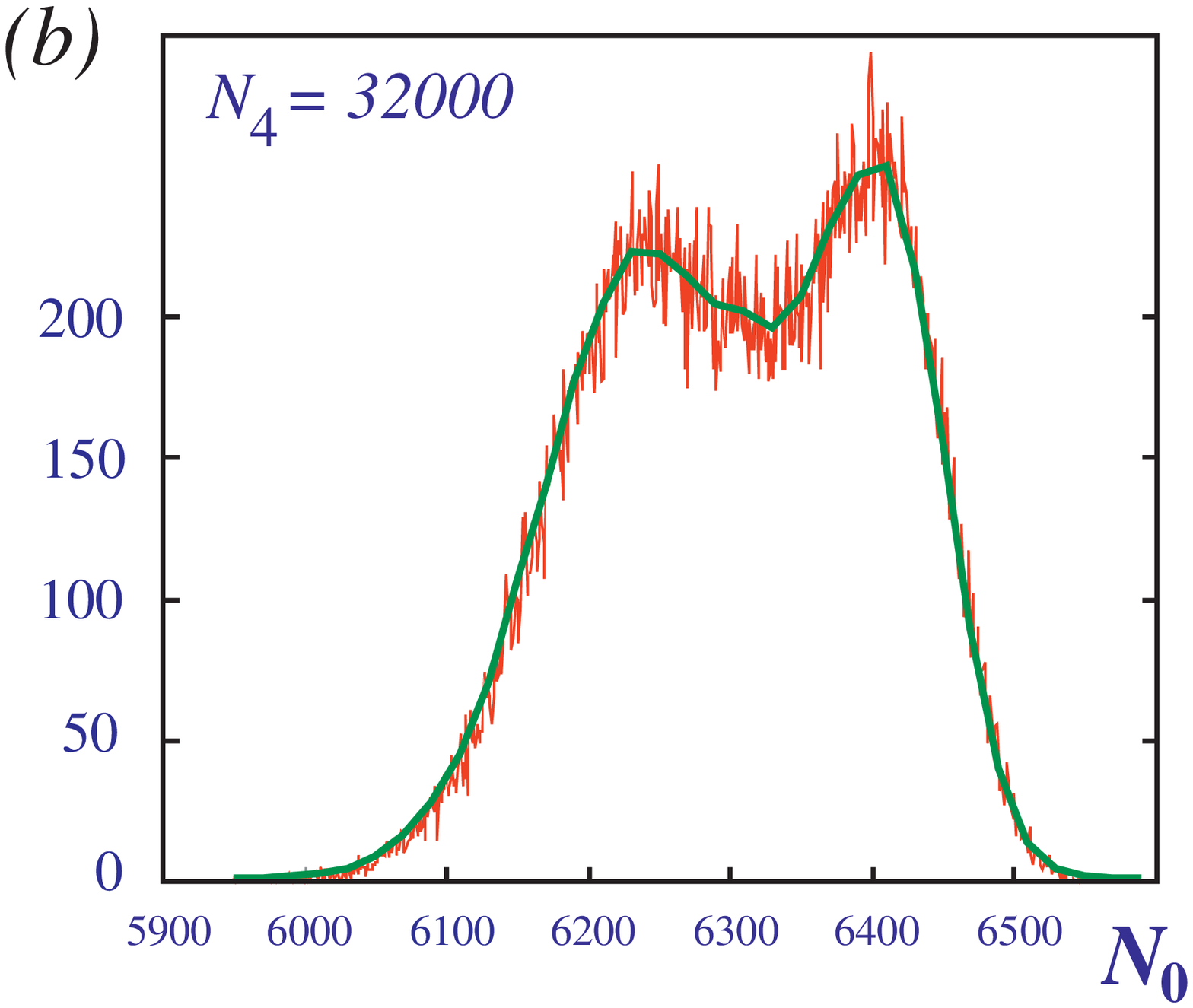,width=3.2in}}
\label{fig.pgenergy}
\caption{ {\small 
  Histogram of the energy density, $n_0 = N_0/N_D$, for
  pure simplicial gravity at the transition $\kappa_c$.
  Both in ({\sl a}) three \cite{jap3d}, and 
  ({\sl b}) four \cite{pet4d} dimensions a two state signal
  is observed, characteristic of a discontinuous phase transition.} 
  $$
  $$
 \hspace{5mm} In four dimensions the transition is weakly first
 order and, consequently, hard to observe  --- 
 it is only manifest on large volumes, $N_4 \gsim 32.000$.
 As present day computers limit practical simulations to 
  $N_4 \lsim 10^5$, this poses a serious
 problem for any numerical investigation of the
 model Eq.~(\ref{part.simplicial}).
 }
\end{figure}
\afterpage{\clearpage}


\section{Modified Models of Simplical Gravity}

\noindent
The discontinuous nature of the phase transition
implies that the model Eq.~(\ref{part.simplicial}) 
does not posses a continuum limit where
a theory of gravity could be defined.
Does this mean that there is something inherently
wrong with the dynamical triangulation approach, 
or is the discretization Eq.~(\ref{part.simplicial})
simply too naive --- some fine-tuning of
additional parameters might be needed.
The success of two-dimensional models of dynamical 
triangulations, where the critical behavior
is remarkable insensitive to details of
the discretization, might be misleading.
The theory of gravity in higher dimensions
is more complicated and more carefully
defined discretization may be necessary.

This has led to investigations of modified models of
simplicial gravity; modified either by changing the relative 
weight of the triangulations with e.g.\ a {\it measure term},
or by coupling {\it matter fields} to the theory.  
Although preliminary investigations of such modified models 
did not observe any qualitative change in the critical 
behavior \cite{3dmatter,measure}, more recent 
numerical simulations \cite{sim3d,jap3d,us4d} 
have revealed that such modifications do indeed
influence the phase structure:
The phase transition appears to soften, 
or even disappear.  Moreover, a new {\it crinkled} phase 
is observed, characterized by internal geometry closer to what 
one might expect of smooth manifolds.

Two such modified models of simplicial gravity
have recently been investigated in $D>2$:

\vspace{6pt}
\noindent
({\tt a}) {\it A modified measure}:
\begin{equation}
 Z_{N_D}(\kappa,\beta) \;=\;
   \sum_{T \in {\cal T}_{N_D}} 
   \prod_{i\in N_{D-2}} \!\! o(n_i)^{\beta} 
    \;\; \frac{{\rm e}^{\;\kappa N_{D-2}}}{C(T)},
 \label{part.measure}
\end{equation}
where $o(n_i)$ is the local volume of an sub-simples
$i$ --- the number of $D$--simplexes containing $i$.

\vspace{6pt}
\noindent
({\tt b}) {\it Coupling to $f$ copies of matter fields}:
\begin{equation}
 Z_{N_D}(\kappa,f) \;=\;
 \sum_{T \in {\cal T}_{N_D}} 
 \frac{{\rm e}^{\; \kappa N_{D-2}}}{C(T)} 
 \; \left ( Z_{\rm M}(T) \right )^f
 \label{part.matter}
\end{equation}
\vspace{-5pt}
where $Z_M$ is the matter partition function.

\vspace{12pt}
In Ref.~\cite{us4d} the particular case of vector fields 
coupled to four-dimensional simplicial
gravity was investigated:
\begin{eqnarray}
 Z_{\rm M}(T) &=  &\int \prod_{l\in T} \; {\rm d}A(l) 
  \; {\rm e}^{ \; \textstyle - S_{\rm VF}}  \\ \nonumber
 S_{\rm VF} &=  &\sum_{t_{abc}} \; o(t_{abc})
         \left [ A(l_{ab})+A(l_{bc})+A(l_{ca}) \right ]^2
 \label{part.vector}
\end{eqnarray}
where $A(l)$ are non-compact $U(1)$ gauge fields residing
on the edges $l$, $A(l_{ab})$~=~ $- A(l_{ba})$,
and the sum is over all triangles $t_{abc}$ in the 
four-dimensional triangulation.

\subsection{Lessons from two dimensions}

Before discussing the phase structure 
of the models Eqs.~(\ref{part.measure}) and (\ref{part.matter})
for $D>2$, it is worth considering the impact the
two modifications have on simplicial gravity 
in two dimensions.\footnote{In 
$D = 2$ the phase diagram simplifies
as the Einstein term in the action is a topological invariant 
and the dynamics is insensitive to the corresponding coupling
constant (as long as the topology of the manifold is fixed).}

In two dimensions modifying the measure has no
effect on the critical behavior, except for 
large negative values of the coupling $\beta$ where
formations of large curvature defects is enhanced and
there is a cross-over to a crumpled phase, dominated by
intrinsically collapsed configurations.

In contrast, adding matter fields  
changes the critical behavior.
For matter fields with central charge $c \leq 1$ 
this corresponds to coupling conformally
invariant matter to $2D$--gravity (Liouville theory).
In this phase the fractal structure displays
a remarkable degree of universality; it only
depends on the total central charge of the 
matter fields: $ -\infty  \leq \gamma_s(c) \leq 0$, and 
$2 \leq d_H(c) \lsim 4$.
For $c > 1$, on the other hand, the
fractal structure degenerates into branched polymers.
This is understood qualitatively as a condensation of {\it spiky}
configurations \cite{spikes};  
the free energy of spikes, $F_{\rm spike} \sim (1-c)\log (1/a)$,
estimated from the dynamics of the conformal factor,
shows that such singular configurations dominate the
partition function for $c > 1$.

\begin{figure}[t]
\centerline{\psfig{figure=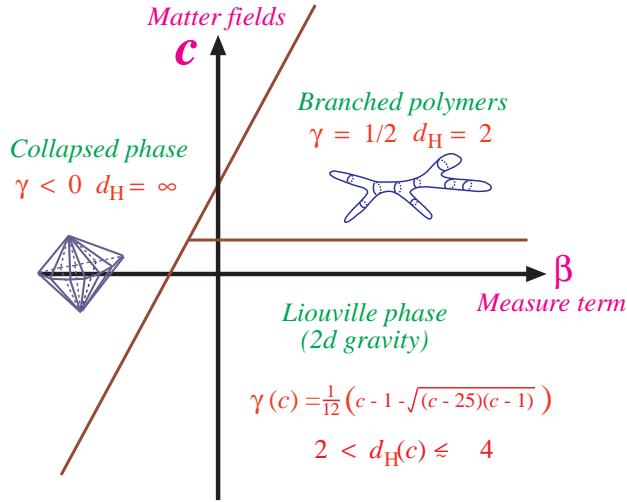,width=3.2in}}
\label{fig:2dphase}
\caption{\small The phase structure of two-dimensional
 simplicial gravity coupled to matter fields and/or 
 a measure term.}
\end{figure}

What does this imply for the phase diagram in higher
dimensions?  Is it possible that the similarity between 
the phase structure for pure gravity in $D>2$ and
the $c>1$ region in Figure~4
is not coincidental. That a suitable tuning
of the parameters $\beta$ or $f$, could lead to some 
higher-dimensional analog of the Liouville phase? 

This scenario is supported by recent analysis
in Ref.~\cite{mottola}
which suggest that something analogous to the ``$c=1$ barrier''
might exist in four dimensions.
Using an effective action for the conformal factor, 
they calculate the free energy of spikes:
\begin{equation}
 F_{\rm spike} \;\sim\; {\textstyle \frac{1}{2}}
  (Q^2 - 8) \; \log \left ( 
  {\textstyle {\frac{1}{a}}}\right ).
\end{equation}
 $Q^2 = {\textstyle \frac{1}{180}}
  \left ( N_S + \frac{\scriptstyle 11}{\scriptstyle 2}N_F
       + 62N_V - 28 \right ) \;+\;  Q^2_{\rm Grav} $
plays the role of a central charge and $N_{S,F,V}$ refers to
the number of massless scalar, fermion and vector fields
contributing to the trace anomaly.  A one-loop 
calculation of the contribution from the transverse gravitons,
$Q^2_{\rm Grav} \approx 7.9$, suggests that in
the absence of matter fields the theory has no sensible
stable vacuum.  It is intriguing,
however, that the contribution of matter and ghost fields
appears in $D=4$ {\it opposite} to that found in $D=2$;
in four dimensions adding matter fields might actually 
stabilize the theory!

\subsection{Modified phase structure in $\mathbf D>2$}

Motivated by these considerations, extensive numerical
investigations have been carried out
on the models Eqs.~(\ref{part.measure}) and
(\ref{part.vector}) \cite{jap3d,sim3d,us4d}.  
In addition to MC simulations,
a method for calculating the strong-coupling
expansion of simplicial gravity in $D>2$ 
has been developed \cite{us4d}, and has proved
invaluable for a qualitative view of the 
impact different modifications have on
the critical behavior.
The results of these investigations, which suggest
a phase diagram akin to Figure~5,
can be summarized as follows:

\begin{figure}[t]
\centerline{\psfig{figure=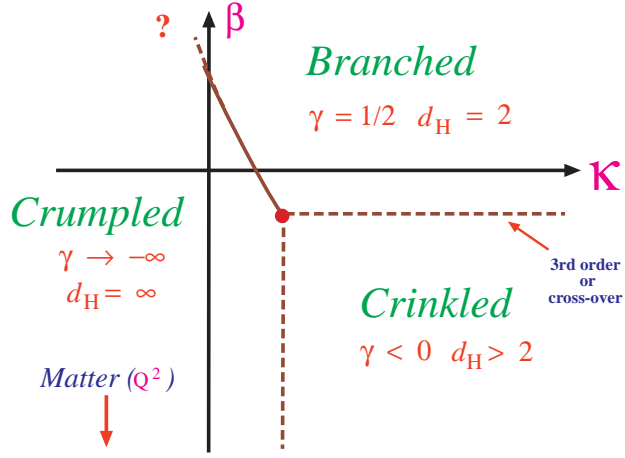,width=3.2in}}
\label{fig:phase4d}
\caption{\small A schematic phase diagram of modified models
 of simplicial gravity for $D >  2$.}
\end{figure}

\begin{mydescription}

\vspace{-3pt} 
\item[({\tt i})]
 As several vector fields are coupled to simplicial gravity
 (or the curvature fluctuation are enhanced by modifying 
 the measure) the discontinuous phase transition (solid line), 
 separating the crumpled and the elongated phases, is
 softened and eventually, at some value $f \gsim 3$ 
 ($\beta \lsim -1$), either becomes continuous or disappears
 altogether.

\vspace{-3pt} 
\item[({\tt ii})]
 At the same value of $f$ ($\beta$) the polymerizations of the
 geometry is suppressed, the elongated phase disappears,
 and a new {\it crinkled} phase appears.  The crinkled
 phase is separated from the crumpled and elongated phases
 by lines of either continuous  
 phase transitions or by a crossover region (dashed~lines).
 
\end{mydescription}

\vspace{-5pt}
In contrast to two dimensions, a remarkable
degree of universality appears in 
simplicial gravity in $D>2$;
all ``reasonable'' modifications
seem to lead to the same qualitative 
phase structure Figure~5 \cite{us4d}.

The nature of the crinkled phase has been
explored for $D =4$,
both in MC simulations and from
the strong-coupling expansion \cite{us4d}.  As the number
of vector fields $f$ is increased (or $\beta$
decreased), the string susceptibility exponent
$\gamma_s$ decreases (Figure~6).
Moreover, modulo some trivial re-scaling: 
$\beta = - {\scriptstyle \frac{1}{2}} f - 
{\scriptstyle \frac{1}{4}}$, the results for 
the two models are identical.
Measurements of the fractal dimension in the crinkled phase
yield $d_H \approx 4$.

At the transition lines separating the
crinkled phase from the rest of the phase
diagram (the dashed lines), 
no divergence is observed in the specific
heat, only a cusp.  It is thus difficult to
determine in MC simulations if these are
soft continuous transitions (third or higher
order) or if this represents a smooth cross-over
to different a fractal structure.  
However, a soft continuous
phase transition would not be unexpected;
coupling to geometric disorder tends to soften phases 
transitions.  For example, most transitions for
matter coupled to $2D$--gravity are of third order.

\begin{figure}[t]
\centerline{\psfig{figure=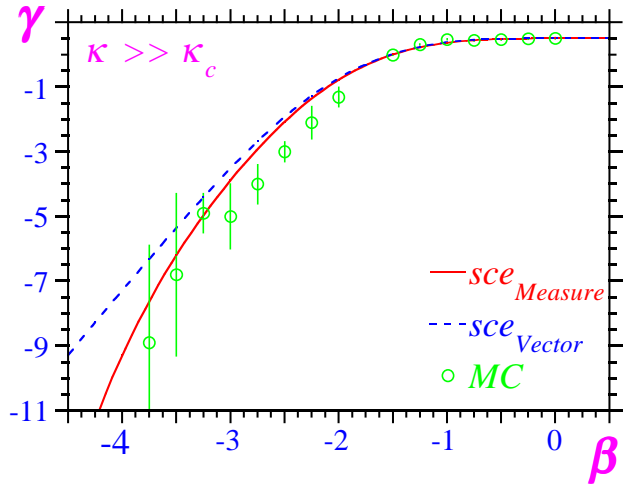,width=3.2in}}
\label{fig:gamma}
\caption{\small The exponent $\gamma_s$ {\it versus} a modified measure 
 (solid line) or coupling to matter fields (dashed)
 from the strong-coupling expansion for $\kappa \gg \kappa_c$. 
 The circles are from MC simulations of the model Eq.~(\ref{part.measure}) 
 for $\kappa = 4.5$ \cite{us4d}.}
\end{figure}

How should one interpret the phase structure Figure~5?
Does the crinkled phase play a role analogous to the Liouville
phase in two dimensions, or is the internal geometry
just as pathological in this phase as in the
crumpled and elongated phases?
Alternatively, is a continuum limit possible
at some of the observed transition lines; 
provided, of course, they correspond continuous 
phase transitions? 
Further investigations will hopefully
shed some light on these issues.

\newpage

\subsection{The strong-coupling expansion}

The strong-coupling expansion --- an
explicit enumeration of all triangulations 
up to a given volume --- provides a powerful tool
for exploring the phase structure of different
modifications of simplicial gravity.
It has previously been used in two dimensions,
where a matrix-model formulation 
allows a recursive construction of all triangulations 
\cite{strong2d}.  
Combined with an appropriate series extrapolation
method, the strong coupling expansion yields,
for some models, very accurate
estimates of the critical exponents.

In $D>2$ a recursive algorithm for generating all 
triangulations is not available.
Instead an alternative procedure constructing
the series in numerical simulations
has been developed \cite{us4d,bengtproc}:

\begin{mydescription}

\vspace{-3pt}
\item[({\tt i})]
 Identify all distinct triangulations $T$ using MC
 simulations and a sufficiently complicated ``hash'' function.

\vspace{-3pt} 
\item[({\tt ii})]
 Calculate the corresponding symmetry factor $C(T)$ by
 comparing all permutations of the vertex labels.

\vspace{-3pt} 
\item[({\tt iii})]
 For each triangulation $T$ calculate the appropriate
 contribution from matter fields and/or a modified measure.
 
\end{mydescription}

\vspace{-4pt}
The hash function $f(T)$ plays a vital role
in the identification.  It should be sufficiently complicated 
to distinguish between combinatorially inequivalent 
triangulations, yet simple enough to be repeatedly calculated.  
The identification is
verified by comparing the calculated symmetry factors 
$C(T)$ to the relative frequency with which different
triangulation appear in the MC simulations.
In practice, the identification procedure is much simplified 
by observing that only triangulations not reducible
by any of the volume decreasing geometric moves have to 
be considered in the MC; all others are 
systematically constructed from smaller ones.
In practice 99.9\% of the triangulations turn out to
be reducible.

This method has been used to calculate the first
15 to 20 terms in the strong-coupling expansion
of the partition function Eq.~(\ref{part.simplicial}),
both in three and four dimensions.
This corresponds to roughly $10^6$ distinct triangulations.
Given the triangulations it is
easy to calculate the contribution from different
types of matter fields, or from a modified measure, to the
partition function.  To determine the critical 
behavior, appropriate series extrapolation methods
are used, e.g.\ the {\it ratio method} \cite{strong2d},
in which higher order corrections
to the partition function are systematically
eliminated.

The method is, however, limited to the region of
the phase-space where the partition function is
rapidly convergent.   In cases where the sub-leading
corrections to the partition function are not analytic,
such as in the presence of a confluent
singularity, the extrapolation fails to converge.
In simplicial gravity the strong-coupling expansion 
has provided valuable informations about the nature of 
the elongated and crinkled phases; however, 
it has been of limited use in the crumpled phase
and in the transitions region.

\subsection{``Balls in boxes''}

Many of the characteristics of the phase structure of 
simplicial gravity, described in the previous sections,
are captured by a simple mean-field model ---
the {\it balls-in-boxes} model \cite{balls1}. 
This model consists of fixed number $N$ of balls
distributed into a variable number $M$ of boxes, 
$1 \leq M \leq M_{\rm max}$.  In its partition function,
\begin{equation}
 Z_N(\kappa,\beta) \;=\;
  \sum_{m=1}^{M_{\rm max}}  \; {\rm e}^{\,\kappa m}
 \sum_{q_1,...,q_m} p(q_1)... p(q_m)
  \;\delta_{q_1+...+q_m,N} \;,
 \label{part.balls}
\end{equation}
the partitions of balls are weighted by the product
of one-box weights $p(q_i) = q_i^{-\beta}$, where $q_i = 1,2,...$
is the number of balls in box $i$.  This particular
choice of weights mimics the modified measure,
Eq.~(\ref{part.measure}), introduced in simplicial gravity.

The model Eq.~(\ref{part.balls}) derives its non-trivial features 
from the constraint on the total number of balls.  
Introducing the average density of balls $\rho = M/N$, the ``curvature'',
there is phase transition at a critical density $\rho_c$.
This transition separates an {\it elongated} ({\it fluid}) phase, 
with the balls evenly distribution among the boxes, 
from a {\it crumpled} phase, where the balls ``condensate''
into one {\it singular} box (akin to Bose-Einstein condensation).
The transition is first order for $\beta \in (2,\infty)$,
whereas it is continuous for $\beta \in (1,2]$.

\begin{figure}[t]
 \centerline{\psfig{figure=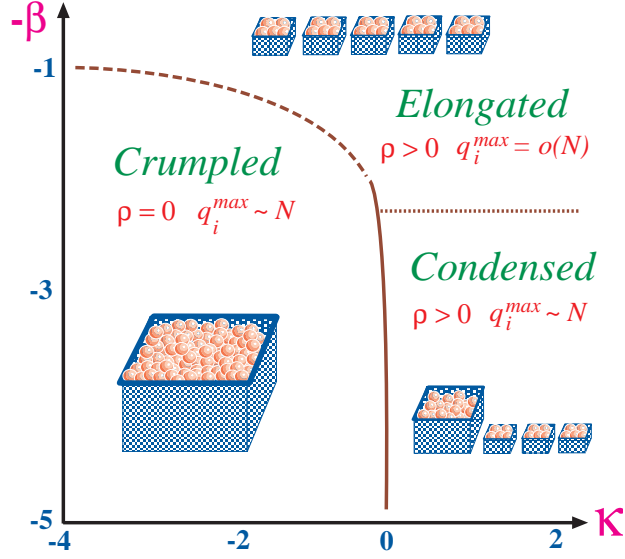,width=3.2in}}
 \label{fig:phaseball}
 \caption{\small The phase diagram of balls-in-boxes.}
\end{figure}
\afterpage{\clearpage}

By introducing an additional kinematic bound on the
average density of balls per box, $\rho < \rho_{art}$, a
new {\it condensed} phase appears in the model
for $\beta \gsim 2.5$ \cite{balls2}.  
This is shown in Figure~7.
In this phase, which resembles  the 
crinkled phase of simplical gravity, the model is 
neither crumpled nor fluid.  The average density per
box is finite, $\langle r \rangle > 0$; however, there is
a singular box in the system which captures a
finite number of the balls: $\langle q_{\rm sing} \rangle /N > 0$.

It is remarkable how such a simple model successfully 
describes many of the geometric features of simplicial 
gravity in $D>2$, e.g.\ the appearance of a singular vertex 
and the phase structure, while ignoring all the 
non-trivial information encoded in the connectivity 
of the triangulations.  


\section{Degenerate Triangulations}

\noindent
In the partition function Eq.~(\ref{part.simplicial}), 
${\cal T}$ denotes a suitable ensemble of
triangulations included in the sum.
Different ensembles are defined by imposing various
restriction on how the simplexes are glued together.
Provided this leads to a well-defined partition function,
and as long as this difference is only at the level
of the discretization, one expects different choices
of ${\cal T}$ to lead to the same continuum theory
in the thermodynamic limit.  
This is, of course, not true if the partition
function is divergent as is the case even
in two dimensions if the
topology is not restricted.
\afterpage{\clearpage}
\begin{figure}[h]
 \centerline{\psfig{figure=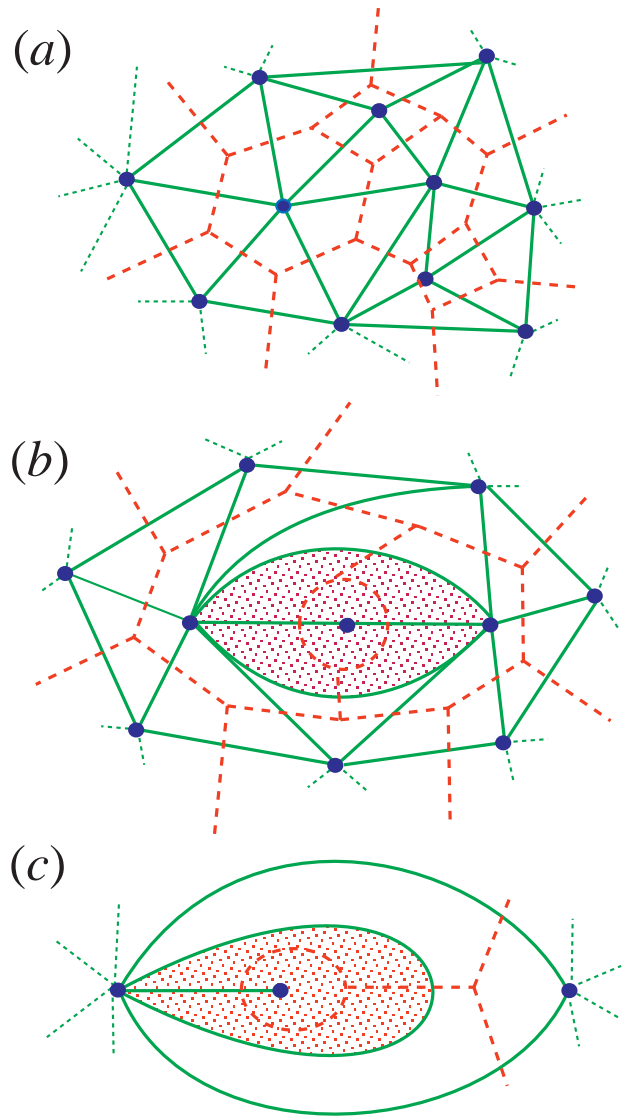,width=3.2in}}
 \label{fig.deg2d}
 \caption{\small Three types of two-dimensional triangulations:
  ({\it a}) combinatorial, ({\it b}) restricted degenerate,
  and ({\it c}) maximally degenerate.  The dashed
  lines indicate the corresponding dual graphs.}
\end{figure}

Traditionally, simulations of simplicial gravity in $D > 2$
have used an ensemble of {\it combinatorial} triangulations
${\cal T}_C$.  In a combinatorial triangulation every
$D$--simplex is uniquely defined by a set of
$D+1$ distinct vertexes --- it is 
combinatorially unique.  This implies
that two distinct simplexes have at most one face in common;
this amounts to exclude both tadpoles and self-energy diagrams
in the corresponding dual graph.  Easing this restriction,
allowing either multiple connected simplexes or simplexes
glued onto themselves, defines ensembles of 
{\it restricted} and {\it maximally degenerate} triangulations,
${\cal T}_{D_R}$ and ${\cal T}_{D_M}$,
respectively \cite{deg3d}.  In the former ensemble to each
simplex there is associated a set of $D+1$ {\it distinct}
vertexes; however, the same set of vertexes may be
shared by any number of distinct simplexes.
In contrast, in a maximally
degenerate triangulation even the set of vertexes associated
to a simplex may be degenerate, i.e.\ it may contain 
the same vertex more than once.
Clearly, ${\cal T}_C \subset {\cal T}_{D_R} \subset {\cal T}_{D_M}$.
Figure~8 shows two-dimensional examples of those 
three different types of triangulations.

In two dimensions the model Eq.~(\ref{part.simplicial})
defined with those different ensembles is solvable as 
matrix model and, in all three cases, it yields 
the same continuum theory \cite{mat2}.
Then why consider different ensembles; why not use the one
most convenient in the simulations?  It turns out that the 
finite-size effects depend strongly 
on the ensemble used \cite{deg2d}.  In particular, the 
finite-size effects are considerable reduced the less 
restrictions are placed on the triangulations.  
An example of this is the volume dependence
of the effective magnetization exponent, $\beta/\nu d_H$, 
for an Ising model coupled to $2D$--gravity
(Figure~9).  This reduction in the finite-size
effects is understandable --- with a large ensemble of
triangulations it is easier to approximate a given
continuum geometry using triangulations of finite volume.

\subsection{$\mathbf D > 2$}

\afterpage{\clearpage}
\begin{figure}[t]
 \centerline{\psfig{figure=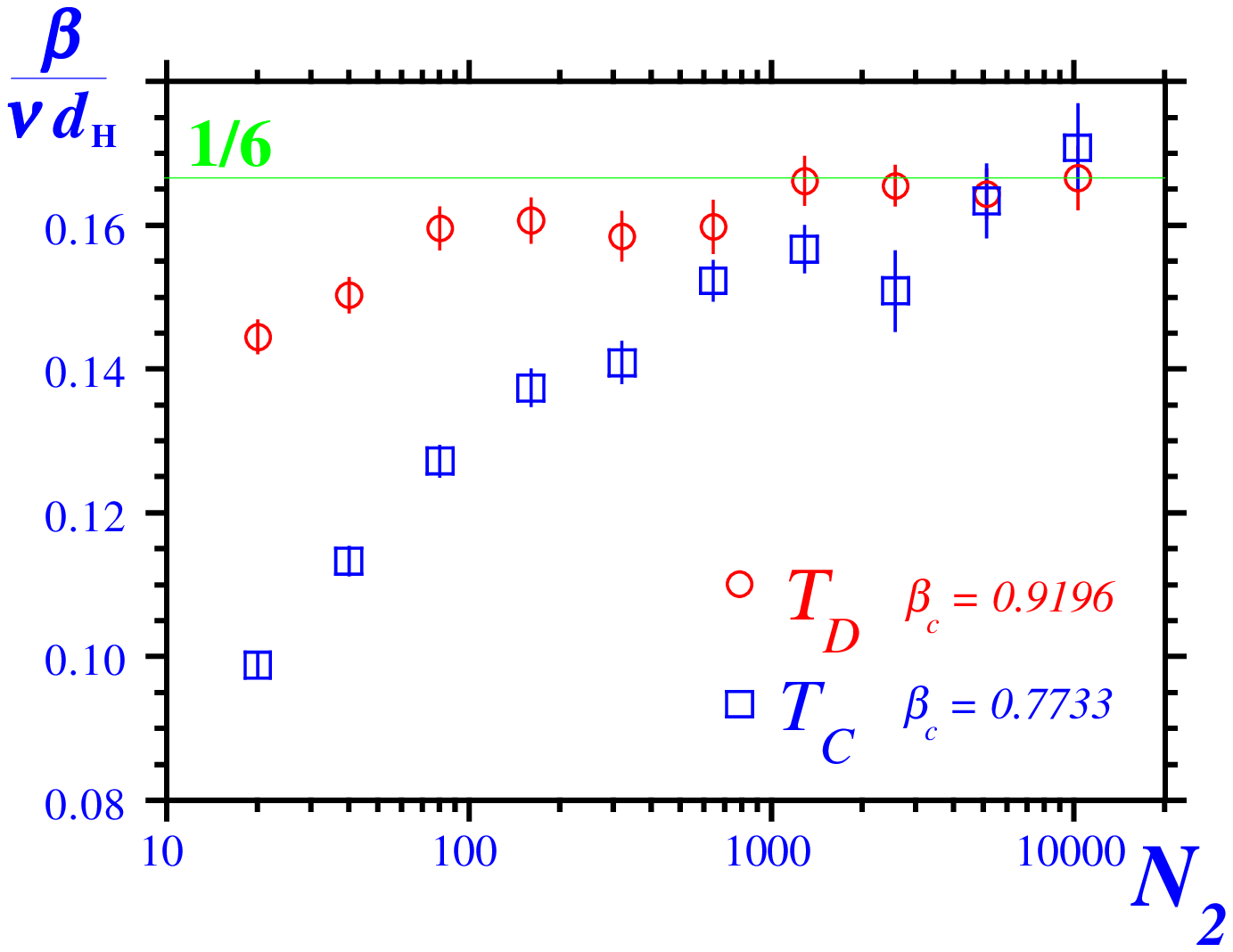,width=3.2in}}
 \label{fig:is2d}
 \caption{\small The effective magnetization exponent, $\beta/\nu d_H$,
  for an Ising model coupled to $2D$--gravity {\it versus} volume 
  \cite{deg2d}, both for combinatorial  (squares) and
  degenerate (circles) triangulations \cite{deg2d}. The theoretic
  value is 1/6.}
\end{figure}

\begin{figure}[tb]
 \centerline{\psfig{figure=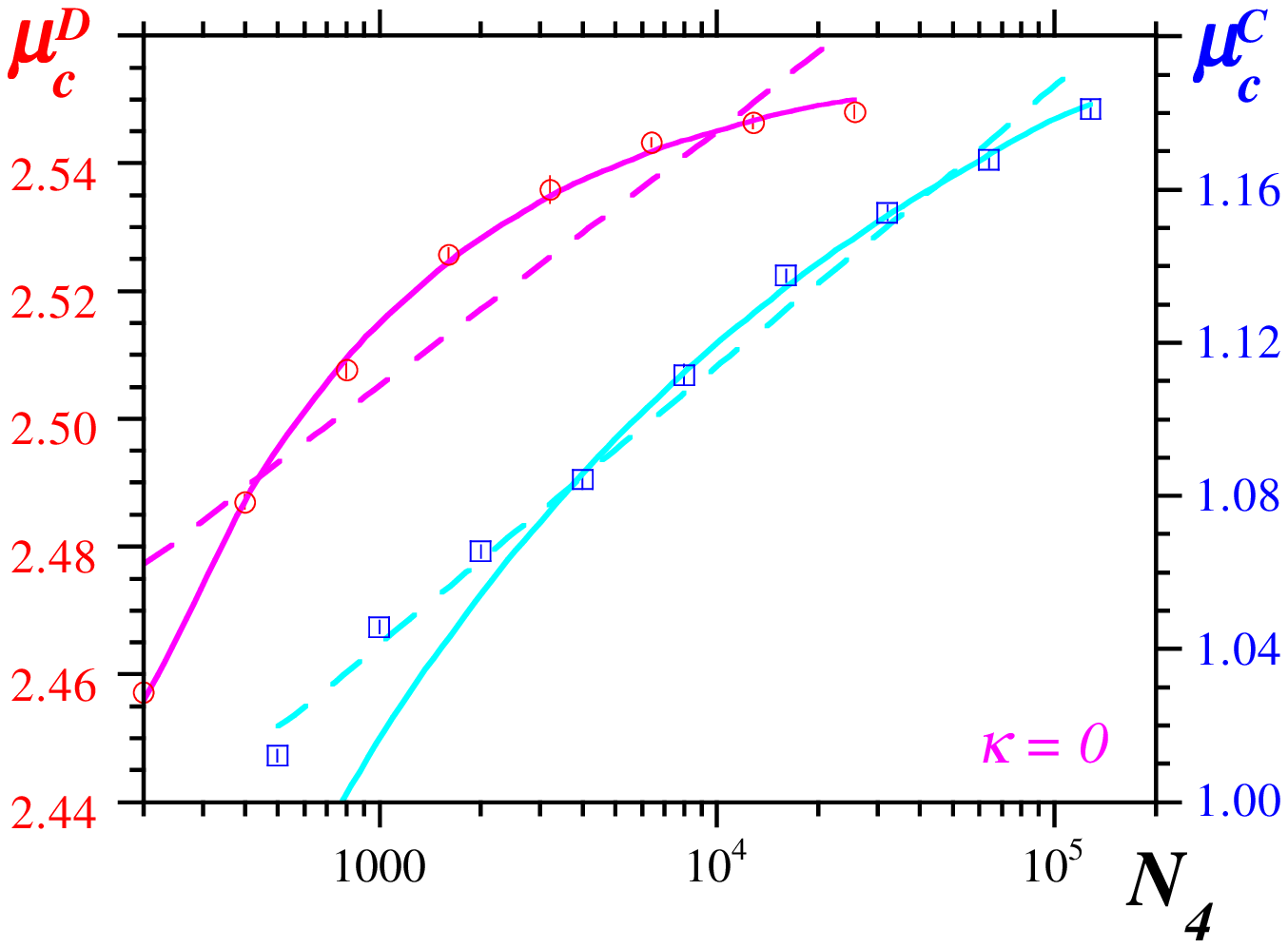,width=3.2in}}
 \label{fig:muc}
 \caption{\small The pseudo-critical cosmological constant
  $\mu_c(N_4)$ for ensembles of $4D$ combinatorial 
  (squares) and degenerate (circles) triangulations.
  Included are fits assuming an power-law convergence (solid lines)
  and a logarithmic divergence (dashed lines) \cite{deg4d}.}
\end{figure}

It is only recently that simulations of simplicial gravity in 
higher dimensions have been extended to include
degenerate triangulations.  In three dimensions both
restricted and maximally degenerate triangulations
have been investigated \cite{deg3d}; however,
only the former ensemble was found to lead to
a well-defined partition function Eq.~(\ref{part.simplicial}).
More recently restricted degenerate triangulations have 
also been used in four dimensions \cite{deg4d}.

Qualitatively, the phase structure does not change when 
degenerate triangulations are included in the 
model Eq.~(\ref{part.simplicial}). 
There is still a small $\kappa$ crumpled phase 
and a large $\kappa$ elongated phase, 
separated by a discontinuous transition.
However, just as in two dimensions the finite-size effects
appear much reduced, especially in the 
crumpled phase.  An example of this is the
pseudo-critical cosmological constant $\mu_c(N_D)$,
which converges much more rapidly for the ensemble ${\cal T}_{D_R}$
than for ${\cal T}_C$.
This is shown in Figure~10 for $D=4$ and $\kappa = 0$.

This important observation demonstrates 
(numerically) an {\it exponential bound} on the number of
degenerate $4D$ triangulations of fixed volume\footnote{
The existence of an exponential bound has been
somewhat controversial for the ensemble of 
combinatorial triangulations in four
dimensions \cite{bound4d}.} ---
a necessary condition for the convergence of the
partition function Eq.~(\ref{part.simplicial}).
This is demonstrated in Figure~10 by two
fits: one fit assumes an exponential bound on the
canonical partition function (solid),
the other a super-exponential growth (dashed).
Whereas for combinatorial triangulations the
fits are somewhat inconclusive, for degenerate triangulations
a super-exponential growth is definitely ruled out \cite{deg4d}.
And, as ${\cal T}_C \subset {\cal T}_D$, this automatically
implies a bound on the entropy of combinatorial 
triangulations as well.


\section{Physical Membranes}

The study of the elastic and geometrical properties of physical
membranes, two--dimensional surfaces with an
extrinsic curvature embedded in ${\rm \bf R}^3$,
is a rapidly advancing field with a nice interplay between
experiments, analytical theory and computer simulations
(see e.g.\ Refs.~\cite{gen.membrane1,gen.membrane2}).
Viewed as two--dimensional generalization of one--dimensional
chains, membranes show a rich behavior on mesoscopic length
scales.  They can be classified according to their internal
structure:
\begin{mydescription}

\vspace{-5pt}
\item[({\tt a})]
 {\it Polymerized} or {\it crystalline membranes} 
 have fixed internal structure
 and are characterized by both long-range orientational and
 translational order.

\vspace{-5pt}
\item[({\tt b})]
 {\it Fluid membranes}, with dynamical
 internal connectivity and no internal order.
\end{mydescription}

\vspace{-5pt}
An example of the former is
the spectrin cytoskeleton of red bloods cells, whereas
bilayers of amphiphiles such as phospholipids generally
exhibit liquid--like behavior.  Physical membranes
encountered in nature are, of course, {\it self-avoiding};
a property pivotal to many aspects of their phase structure.
They are, however, commonly studied as self-intersecting
({\it phantom}) membranes, both for simplicity and as
statistical system interesting in their own right.

It is by now well established, both by analytical and
computational work, that crystalline membranes exhibit a
low--temperature flat (ordered) phase and, at least for
{\it phantom} membranes, a high--temperature
crumpled phase \cite{uscryst}.  The existence of an ordered phase in a
two--dimensional system with a continuous symmetry and
an (apparent) local interaction is remarkable given the
Mermin--Wagner theorem.  What stabilizes the flat phase
in a crystalline membrane are the out--of--plane fluctuations
that couple to the in--plane ``phonon'' degrees of freedom
--- bending of the membrane is accompanied by internal
stretching.

The same argument for a stable flat phase does not apply in
the case of a fluid membrane where the internal
stretching can be compensated by a flow of ``particles'' into
the distorted area, thereby ``screening'' the curvature
fluctuations from the rest of the membrane.  The absence
of a stable flat phase is supported by
renormalization group analysis of models of fluid membranes,
which suggest that at large distances the bending rigidity
becomes irrelevant \cite{rg}.
This is, however, contradicted by numerous numerical simulations
of fluid membranes that indicate a crumpling transition,
akin to that of its crystalline counterpart \cite{fluid}.

\subsection{Including anisotropy}

The above discussion applies to {\it isotropic} membranes.
Remarkable, adding {\it intrinsic anisotropy} 
to a model of crystalline
membrane profoundly influences
the global phase diagram \cite{rt}.  
Most notably, a new {\it tubular} phase appears,
characterized by the presence of
long-range orientational order in one direction only ---
in the transverse directions the membrane is crumpled.  
For a phantom membrane, the tubular phase separates
the existing crumpled and flat phases;
only an isotropic membrane passes directly from a flat 
to a crumpled state (Figure~11).
Analogous to a flat membrane, a tubular membrane 
is stabilized by the transverse stretching energy cost of bending
fluctuations in the extended direction.

\begin{figure}[t]
 \centerline{\psfig{figure=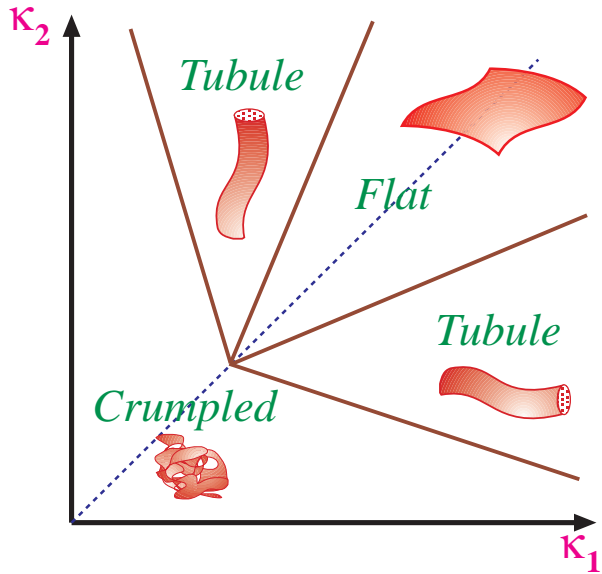,width=3.0in}}
 \label{fig.ancryst}
 \caption{\small A schematic phase diagram of a phantom anisotropic 
  crystalline membrane.}
\end{figure}

The existence of a tubular phase has been confirmed
in recent numerical simulations\footnote{
Simulations of physical membranes are notoriously
difficult due to very long auto-correlation times present
in standard MC simulations.  These auto-correlations are, though,
much reduced by applying more powerful updating methods
such as overrelaxation or unigrid algorithms in the
simulations \cite{algo}.}
on a crystalline membranes \cite{ustub}.  The membrane is discretized
on a triangular mesh, with the topology of a disk, and
the interactions modeled by Hamiltonian composed of
two terms: a Gaussian pair potential between neighboring
nodes, and a bending energy introduced as 
ferromagnetic interaction between normals ${\bf n}_a$
on adjacent triangles:
\begin{equation}
 {\cal H}[{\rm \bf r}] \;=\; \sum_{\langle\sigma \sigma^\prime\rangle} 
 \left|
 {\rm \bf r}_{\sigma}
 - {\rm \bf r}_{\sigma^{\prime}} \right|^2
 \; - \; \kappa_1 {\sum_{\langle ab\rangle}}^{(x)} 
  {\bf n}_a\cdot{\bf n}_b \;-\; \kappa_2 {\sum_{\langle ab \rangle}}^{(y)} 
  {\bf n}_a\cdot{\bf n}_b \,.
 \label{part.tubule}
\end{equation}
The anisotropy is introduced as different bending rigidity
in the two intrinsic directions, $x$ and $y$.

The phase diagram of the model Eq.~(\ref{part.tubule}) was
explored, for different bending rigidities ($\kappa_1\,$,$\,\kappa_2$),
and evidence for two distinct phase transition ---
a crumpled--to--tubular and tubular--to--flat ---
is observed in the  fluctuations in the two bending
energy terms: $C_V^x(\kappa)$ and $C_V^y(\kappa)$
(Figure~12).

In the tubular phase,
the tubule cross-section radius $R_{\perp}^G$
(the radius of gyration) and the undulations $h_{\rm rms}$ 
transverse to its average axis of orientation, scale like:
$R_{\perp}^G \sim L^{\nu_F}$ and 
$h_{\rm rms} \sim L^{\zeta}$, where $\nu_F$ and $\zeta$ are the
size (Flory) and roughness exponents.
MC simulations of a tubular membrane give: $\nu_F = 0.269(7)$
and $\zeta = 0.850(40)$ \cite{ustub}, compared to
the theoretical predictions: 
$\nu_F = {\scriptsize \frac{1}{4}}$ and $\zeta = 1$ \cite{rt}. 
Notice that $\zeta = 1$ implies that a phantom tubule is only
marginally stable with respect to 
the fluctuations in the transverse directions;
the tubule almost crumples.

\begin{figure}[t]
 \centerline{\psfig{figure=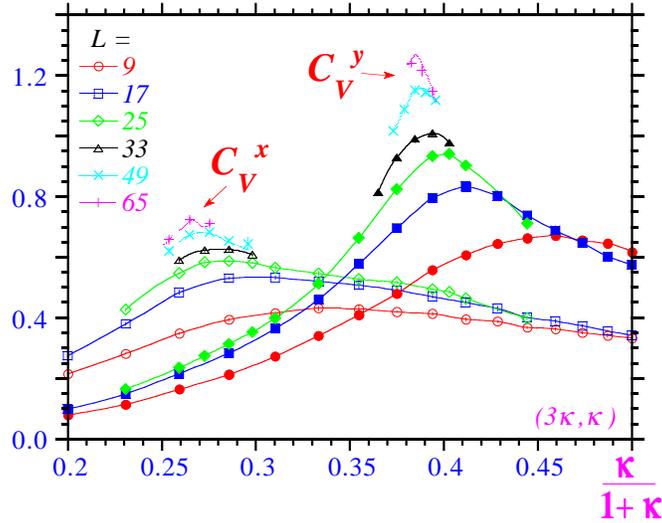,width=3.4in}}
 \label{fig.ancryst}
 \caption{\small The fluctuations in the two bending energy
  terms: $C_V^x$ and $C_V^y$, for a phantom anisotropic 
  crystalline membrane, {\it versus} 
  $(\kappa_1,\kappa_2) = (3\kappa,\kappa)$ \cite{ustub}.}
\end{figure}

\subsection{Self-avoiding membranes}

The more physical self-avoiding membranes present more
of a challenge, especially for numerical simulations,
due to the non-local nature of the self-avoiding constraint.
For a flat isotropic crystalline membrane self-avoidance is expected to
be irrelevant; self-avoidance is, however, expected to suppress
the crumpled phase.  Hence only flat self-avoiding isotropic
membranes are expected to exist;
this is supported by numerical simulations.

For membranes crumpled in one direction only, 
self-avoidance is less constraining
and the tubular phase is expected to survive in the
more physical self-avoiding case \cite{rt,mg,mt}.
However, unlike for a flat membrane self-avoidance is expected to
be relevant for tubules.  Adding a non-vanishing self-avoiding
coupling leads to a flow in the couplings
to a new critical (infrared stable) fixed point, distinct
from the one that governs the critical behavior
of a phantom tubule.  This fixed point structure is 
supported by renormalization group analysis \cite{rt,mt}. 
An improved one-loop calculations of the critical exponents 
$\nu$ and $\zeta$, in an $\epsilon$--expansion around 
the upper critical dimension $d_c^{up}=11$, gives: 
$\nu = 0.62$ and $\zeta = 0.8$ \cite{mt}.

It is of great interest, albeit very challenging,
to verify this critical behavior in numerical simulations
of a self-avoiding anisotropic membrane.  
This work is underway \cite{proc.mark}. 
Preliminary results indicate the existence of only 
one phase transition, from a tubular to a flat phase; as
expected a crumpled phase is not observed.


\section{Outlook}

\noindent
({\tt i})  Much work is needed before the phase structure
of modified models of simplicial gravity in $D >2$
is fully understood.  The nature of the 
crinkled phase, and of the observed 
phase transitions, must be established before any statements
can be made about the relevance the observed phase 
structure Figure~5 has for any potential  
continuum theory of quantum gravity. 

\vspace{5pt}
\noindent
({\tt ii})  Degenerate triangulations might prove a valuable
tool in any such investigations.  In addition to 
reducing the finite-size effects in simulations, 
they also serve as a consistency check on any observed phase 
structure in simplicial gravity --- any critical behavior that
is relevant in a continuum limit should be independent
of the ensemble of triangulations used in defining
the model.

\vspace{5pt}
\noindent
({\tt iii})  Anisotropic crystalline membrane have a
surprisingly rich phase structure, some of which has
been revealed through analytic calculations and/or numerical
simulations.  To extend these simulations to the more
physical self-avoiding case is of much interest, albeit
very challenging.

Fluid membranes present another challenge to numerical
simulations.  They are expected always to crumple on long
length-scales, yet computer simulations suggest a 
possible transition to a flat phase.  However, these results where
obtained is simulations of systems of modest size;
revisiting fluid membranes now that better algorithms and faster 
computers are available would be an interesting project.

\end{document}